\documentstyle[twoside,fleqn,espcrc2,epsfig]{article}
\def\`#1{\if#1i{\accent18 \i}\else{\accent18 #1}\fi}
\def\'#1{\if#1i{\accent19 \i}\else{\accent19 #1}\fi}

\newcommand{\AmS}{{\protect\the\textfont2
  A\kern-.1667em\lower.5ex\hbox{M}\kern-.125emS}}

\title{Muon  Bremsstrahlung and Muonic Pair Production in Air Showers.
}
\author{A.~ N.~Cillis\address{Laboratorio de F\'isica Te\'orica\\
Departamento de F\'isica\\
Universidad Nacional de La Plata\\
C. C. 67 - 1900 La Plata\\
Argentina}
and S.~J.~Sciutto$^{\rm a}$\thanks{Fellow of CONICET (Argentina)}}

\begin{document}

\begin{abstract}
The objective of this work is to report on  the modifications in air
shower development due to muon bremsstrahlung and muonic pair
production.  In order to do that we have  implemented new muon
bremsstrahlung and muonic pair production procedures in the AIRES air
shower simulation system, and have used it to simulate ultra high
energy showers in different conditions.

The influence of the mentioned processes in the global  development of
the air shower is important for primary particles of large zenith
angles, while they do not introduce significant changes in the
position  of the shower maximum. 

\end{abstract}


\maketitle

\section{Introduction}

The physics of ultra high energy cosmic rays plays an important
role in our days. 
These  primary particles can not be detected directly. When an ultra
high energy astroparticle interacts with an atom of the Earth's
atmosphere, it produces a number of secondary particles that continue
interacting and generating more secondary
particles. This process is generally known as the air shower.

We have developed a set of programs to simulate air showers and manage
all the output data. Such simulating system is known as
AIRES(AIRshower Extended Simulations)\cite{Aires}.

We started working on the topic of the electromagnetic processes in
air showers some years ago. We analyzed the modification in the
shower development  due to the  reduction of
the electron bremsstrahlung and electron pair production by the LPM
effect and the dielectric suppression \cite{PRD}. We also studied the
influence of the geomagnetic field in an air shower \cite{GF}. 

The main goal of this work is to analyze two  processes that take
place in an  air shower: muon bremsstrahlung and muonic pair production
(electron and positron).
At energies  high enough, these processes become more
important than ionization, and therefore these mechanisms account for
virtually all the energy losses for high energy muons.

In order to study the modifications that these effects introduce in an
air shower, we developed new procedures for such mechanisms
and incorporated them in the AIRES air shower simulation system
\cite{Aires}. The AIRES code has then been used as a realistic air
shower simulator to generate the data used to make our analysis.
  
This work is organized as follows: we start in section 2 with a brief
summary of the theory of the muon bremsstrahlung and muonic pair
production;  in section 3 we show the results of our simulations;
finally  we present our analysis and comments in the conclusion section.

\section{Theory }

\subsection{Theory of the Muon Bremsstrahlung} 

The cross section for muon bremsstrahlung (MB) is calculated by the
standard method of QED \cite{Greiner} similary as in the case of  electron bremsstrahlung. 

The first approach to the MB theory  was due to Bethe and Heitler
\cite{BH,BH2,Rossi}. They considered 
in their calculation the screening of the atomic electrons. 

After this first formulation some  corrections were introduced. Kelner,
Kokoulin and Petrukhin \cite{KKP} also  included the bremsstrahlung
with the atomic electrons. 
The nuclear form factor was investigated by Christy and Kusaka
\cite{CK} for the first time and then by Erlykin \cite{E}. Petrukhin
and Shestakov \cite{PSH} found that the influence of the nuclear
form factor is more important than the ones predicted by the previous
papers. This last
results have been confirmed by  Andreev et al. \cite{ABB}
who also considered the excitation of the nucleus. 

\subsection{Theory of Muonic  Pair Production}

In the lowest significant order of perturbation theory the muonic pair
production(MPP) is a process of 4th order in QED.

Racah \cite{R} was the first to calculate the MPP cross section in
the relativistic region without taking into accuont the atomic and the
nuclear form factor.  Thereafter, Kelner \cite{K} included the
correction due to the screening of the atomic electrons.
The analytical expression for any degree of screening was introduced
by Kokoulin and Petrukhin \cite{KP}.
Those authors also took into account  \cite{KP2} the correction due
to the nuclear form factor. We wish to emphasize that the
influence of the nuclear size is more important when  the energy
tranferred to the pair is large \cite{KP2}. This last case is
important for the air showers and therefore the nuclear size effect
needs to be taken into account in the simulations. 

\section{Air Showers Simulations}

We have calculated the total cross section, and equivalently, the mean free
path (MFP), for  both MB and MPP processes.

\begin{figure}[htb]
\epsfig{file=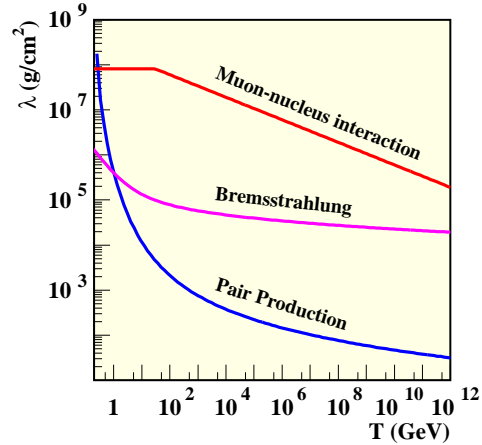,width=60mm}
\caption{Mean free path: MB, MPP and muon nuclear interaction vs the
kinetic energy of the initial muon.}
\label{fig:mfp}
\end{figure}

Figure \ref{fig:mfp} shows the MFP  in  $g/cm^2$ for MB, MPP
and muon-nucleon interaction as functions of the  kinetic energy of the
initial muon. 
The information in this figure can be compared  with the depth of the
atmosphere. The vertical depth of the atmosphere is about 1000
$g/cm^2$ while in the case of very inclined showers ($85^\circ$  of
zenith  angle) is about 9000  $g/cm^2$. Therefore, the probabilities
of  MB and MPP will not be very large except in the case of large
zenith angles.
The muonic component of the showers (number of muons divided by the
number of electrons and gammas) at ground level becomes very
important for zenith angles larger than $60^\circ$ (see, for
example, figure 2 in reference \cite{GF}); this is another reason
to expect that the influence of both effects will be more
appreciable under those conditions.

The MFP's  for MB and MPP diminish when the
kinetic energy of the muon increases and therefore, the influence of
both process will be more important for large energies.
 
Due to the fact that the MFP of muon-nucleus interaction \footnote{We
used the same parametrization of the total cross section used by
GEANT\cite{Geant}} is larger (more than one order of magnitude) than
the corresponding for MB we do not take this effect into account in
our calculations.

In order to analyze the influence of MB and MPP in the development
of  air showers initiated by ultra high astroparticles
we have  performed simulations using the AIRES program \cite{Aires}
with different initial conditions: primary particles (protons, irons,
muons), primary energies ($10^{18}$, $10^{19}$ and $3\times 10^{20}$ eV) and
zenith angles.

\begin{figure}[htb]
\epsfig{file=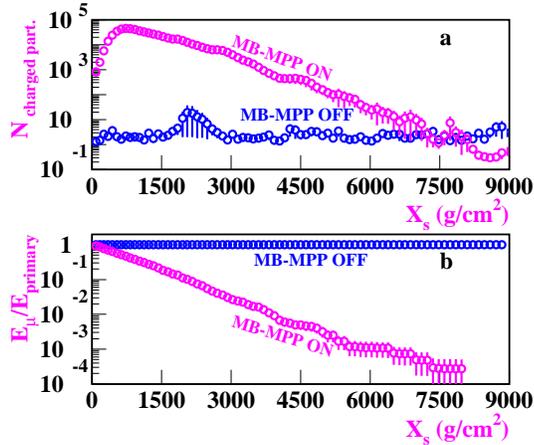,width=70mm}
\caption{a)Longitudinal development of all charged particles vs the slant
depth. Primary particle (Energy): muon ($10^{14}$ eV). b)Longitudinal
development of muon fractional energy vs the slant depth.}
\label{fig:mld}
\end{figure}

To start with the analysis of the simulated data, let us consider the
case of a single muon (eventually produced during the development of a
given shower). This particle may generate a secondary shower if the
processes of MB and MPP are taken into account. This effect is clearly
illustrated in figure \ref{fig:mld}a where the longitudinal
development of all
charged particles is plotted versus the slant depth. In this case the
initial particle is a muon of  $10^{14}$ eV.
When MB and MPP  are not taken into account there is practically no
production of new particles. On the other hand, when the effects are
considered, it is possible to appreciate a secondary shower.

In figure \ref{fig:mld}b the longitudinal development of the muon
fractional energy (muon energy/primary energy) is plotted.
When the effects are not taken into account the muon practically does not
loss energy during all its path, while if the effects are considered,
the muon energy loss is significant.
%
%

We have also studied the modifications in the global observables of
the shower, for example when the primary particle is a proton.

\begin{figure}[htb]
\epsfig{file=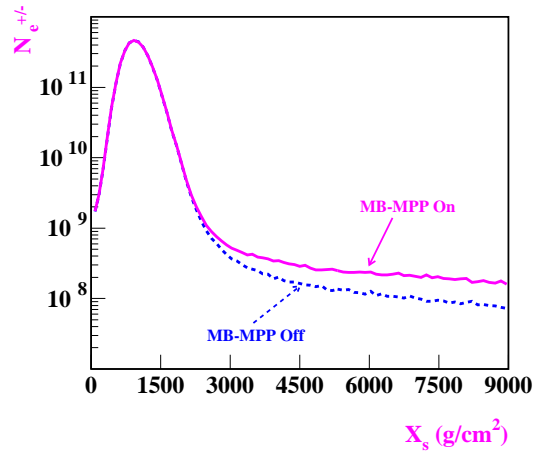,width=70mm}
\caption{Longitudinal development of electrons and positrons vs the
slant dept. Primary particle (Energy) proton ($3\times 10^{20}$ eV).}
\label{fig:pldep}
\end{figure}

\begin{figure}[htb]
\epsfig{file=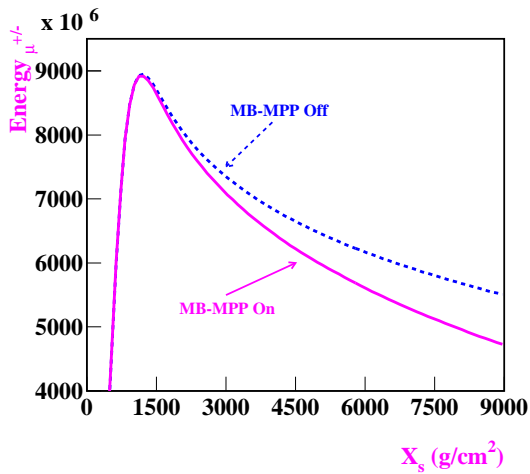,width=70mm}
\caption{Longitudinal development in energy of muon vs the slant depth. Primary
particle (Energy) proton ($3\times 10^{20}$ eV).}
\label{fig:pldm}
\end{figure}

We have plotted in figures \ref{fig:pldep} and \ref{fig:pldm} different
obsevables for $3\times 10^{20}$ eV proton shower with zenith angle $85
^\circ$, taking and not taking into a account the effects
of MB and MPP. Figure \ref{fig:pldep} represents the longitudinal
development  of electrons and positrons ($e^{+/-}$) versus the slant depth. The
number of $e^{+/-}$ increase when the effects are taken into
account. The differences between the two cases are
clearly noticeable in the tail of the showers.
The longitudinal development in energy of muons is plotted in
figure \ref{fig:pldm}. In this case the difference is about 15\% at
the tail. The energy of muons diminishes when the effects are
considered.

For the longitudinal development of
muons versus the slant depth the difference  between the cases taking and
not taking into account the effects is less significant, about 2 \%
(not plotted here).

The modification in the shower
development due to the MB and MPP diminishes for small zenith angle of
the primary particle.  We did not find significant modification of the shower
development for zenith angles smaller than $45^\circ$.

\section{Conclusions}

Due to MB and MPP, a high energy muon in the shower can generate
a  secondary shower. This practically does not  occur if the
mentioned mechanisms are not taken into account. 

The influence of MB and MPP in the total development of the air showers is
more important for primary particle of large zenith angles. Below
zenith angles of 45 degrees we do not observe any significant
difference between the cases where these effects are or not taken
into account. 

The MB and MPP do not generate visible modifications in the position of
the maximum development of the shower.
The modification that we observe affect the shower development well
past its maximum (tail of the shower).

We are performing more simulations in order to make a complete study
of the influence of MB and MPP in an air shower \cite{CS}.

\section{Acknowledgments}

This work  was partially supported by CONICET, Agencia Nacional de
Programaci\'on Cient\'ifica and FOMEC program, Argentina.



%

\end{document}